# Powder Diffraction Crystal Structure Determination Using Generative Models

Qi Li[1,2,‡], Rui Jiao[3,4,‡], Liming Wu[5,6], Tiannian Zhu[1,2], Wenbing Huang[5,6,*], Shifeng Jin[1,2,*], Yang Liu[3,4], Hongming Weng[1,2], Xiaolong Chen[1,2]

1 *The Beijing National Laboratory for Condensed Matter Physics, Institute of Physics, Chinese Academy of Sciences, Beijing 100190, China*

2 *School of Physical Sciences, University of Chinese Academy of Sciences, Beijing, 101408, China.*

3 *Dept. of Comp. Sci. & Tech., Institute for AI, Tsinghua University*

4 *Institute for AIR, Tsinghua University*

5 *Gaoling School of Artificial Intelligence, Renmin University of China*

6 *Beijing Key Laboratory of Big Data Management and Analysis Methods, Beijing, China*

‡ Those authors contribute equally.

* Corresponding author: hwenbing@ruc.edu.cn, shifengjin@iphy.ac.cn

**Abstract**

Accurate crystal structure determination is critical across all scientific disciplines involving crystalline materials. However, solving and refining inorganic crystal structures from powder X-ray diffraction (PXRD) data is traditionally a labor-intensive and time-consuming process that demands substantial expertise. In this work, we introduce PXRDGen, an end-to-end neural network that determines crystal structures by learning joint structural distributions from experimentally stable crystals and their PXRD, producing atomically accurate structures refined through PXRD data. PXRDGen integrates a pretrained XRD encoder, a diffusion/flow-based structure generator, and a Rietveld refinement module, enabling the solution of structures with unparalleled accuracy in a matter of seconds. Evaluation on MP-20 inorganic dataset reveals a remarkable matching rate of 82% (1 sample) and 96% (20 samples) for valid compounds, with Root Mean Square Error (RMSE) approaching the precision limits of Rietveld refinement. PXRDGen effectively tackles key challenges in XRD, such as the precise localization of light atoms, differentiation of neighboring elements, and resolution of overlapping peaks. Overall, PXRDGen marks a significant advancement in the automated determination of crystal structures from powder diffraction data.

# I. INTRODUCTION

Crystal structures provide critical insights into the properties and behaviors of materials. The accurate determination of crystal structures is therefore a cornerstone of many scientific fields, including materials science, chemistry, physics, and geology, driving advancements from drug design to the development of new electronic devices[1,2]. While single-crystal X-ray diffraction (SCXRD) is the standard method for crystal structure determination, many materials are only available in powder form. Powder X-ray diffraction (PXRD) is widely used to characterize metals, minerals, catalysts and ceramics due to its cost-effectiveness and broad applicability[3]. However, solving and refining unknown crystal structures from PXRD data is notoriously challenging. For instance, over 150,000 entries in the Powder Diffraction File[4] have some unresolved coordinates of atoms, highlighting the need for improved methodologies.

In a diffraction pattern, the positions and relative intensities of the diffraction peaks depend on the lattice parameters and the distribution of individual atoms, respectively. PXRD can accurately determine the unit cell, but the overlap of peaks with adjacent diffraction angles causes relative intensities ambiguous, hindering the determination of unknown crystal structures[5]. Consequently, sophisticated technologies like free electron lasers[6], powder electron diffraction[7], and maximum entropy algorithms[8] have been proposed to separate overlapped intensities. Various global optimization algorithms based on direct space approaches, such as genetic algorithms and particle swarm optimization, have been used to deduce the atomic positions[9–11]. However, these methods have no intuition of a stable structure, often require the knowledge of space group and structural units to reduce the free parameters as much as possible. High-throughput DFT search frameworks like FPASS[12] and AXS[13] have been proposed to avoid coarse and faulty candidate structures, but they are computationally intensive and validated on a limited number of datasets with known space groups. Final Rietveld refinements[14] also demand significant human participation and intuition, relying on good initial values for the target structure. Despite significant advancements, structure determination from PXRD remains labor-intensive, time-consuming, and requires rich expertise.

Recent developments in machine learning and artificial intelligence (AI) offer promising new approaches to these challenges[15]. Machine learning has been proven effective in predicting the space group symmetry[16–19], phase[20–22], and unit cell parameters[23,24] from PXRD patterns. Deep learning, in particular, has shown remarkable success in crystal structure prediction (CSP), which aims to obtain the three-dimensional structure of crystals based solely on their composition[25–27]. By introducing the periodicity and symmetry constrains into the network, generative deep learning models have demonstrated impressive structure generation capabilities. However, these CSP-generated structures often suffer from randomness, limited accuracy, challenges in predicting polymorphism, and a lack of experimental validation. Incorporating PXRD patterns through multimodal learning[28] is expected to make CSP models more accurate, robust, and generalizable in predicting stable crystal structures that align with experimental PXRD data. Recently,

CrystalNet[29] and XtalNet[30] have been developed towards end-to-end structure solutions from PXRD using generative models. However, CrystalNet has been validated only on cubic and trigonal systems based on similarities of charge density, while XtalNet has focused on artificial organic structures generated by substituting hundreds of structure units. In addition, PXRDNet[31] is proposed to solve stable crystal structures, but its match rate remains limited. None of these models have yet demonstrated a strong ability to generate experimentally stable structures refined by PXRD data.

In this context, we introduce PXRDGen, an advanced end-to-end neural network designed to significantly enhance the determination of crystal structures from PXRD data. PXRDGen learns structural distributions from both experimentally stable crystals and their corresponding PXRD data, enabling it to generate highly accurate atomic structures validated by Rietveld analysis. The network features an optimized XRD encoder, pretrained through contrastive learning to align PXRD patterns with crystal structures, and a conditional lattice and structure generation network based on diffusion or flow models. Together, these components allow PXRDGen to solve structures with unprecedented accuracy in just a few seconds. PXRDGen's robust sense of stable structures also enables the network to effectively tackle major challenges in XRD, such as accurately determining structures involving light elements like hydrogen or lithium and distinguishing between neighboring elements. Our evaluation of PXRDGen on the MP-20 inorganic dataset demonstrates a high success rate, achieving 82% accuracy with a single candidate and 96% with 20 candidates for valid compounds, corresponding to 75% and 88% accuracy for all compounds, respectively. Furthermore, the RMSE of the generated structures is comparable to the precision limits of traditional Rietveld refinement. These results highlight the potential of PXRDGen to automate and elevate the process of crystal structure determination, delivering significant improvements in both speed and accuracy.

## II. Results and discussion

### 2.1 Overview of PXRDGen

Crystal structure determination from diffraction data typically involves three steps: 1) unit cell determination, 2) structure solution, and 3) structure refinement. In powder X-ray diffraction, unit cell parameters ($L$) can be accurately determined using indexing software, but solving and refining the fraction coordinates ($F_i$) of crystal structure remains the most challenging step. PXRDGen addresses these challenges by automatically solving and refining crystal structures using generative models conditioned on both the chemical formula and normalized PXRD data. As shown in Figure 1, PXRDGen comprises three key modules: the pre-trained XRD encoder (PXE) module in purple, the crystal structure generation (CSG) module in orange, and the Rietveld refinement (RR) module in green. These modules work together to create the first PXRD structure determination neural network capable of generating atomically accurate structures refined by Rietveld methods.

PXRDGen utilizes highly flexible neural networks to optimize its performance. The PXE module employs contrastive learning to align the latent space of PXRD patterns with crystal structures, providing crucial information for generating conditional lattice parameters ($L$) and fraction coordinates ($F_i$). In addition, two distinct PXRD encoders are designed to extract diffraction features: one is based on a convolutional neural network (CNN)[32] and the other leverages Transformer (T)[33] architecture.

The CSG module generates crystal structures by conditioning on PXRD features and chemical formulas, using either diffusion[34] or flow[35] generative frameworks. The main framework and the diffusion model of PXRDGen are adapted from DiffCSP[26] and the flow model is inspired by FlowMM[27]. The flow-based CSG module achieves state-of-the-art match rate and speed in solving stable crystal structures. Unlike traditional Crystal Structure Prediction (CSP), which predicts a compound's stable 3D structure based solely on its composition, PXRDGen allows cell parameters ($L$) to be extracted directly from PXRD data through the Conditional Cell Generation network (CellNet) or conventional indexing results, enhancing the accuracy of the fractional coordinates ($F_i$). The structures generated by the CSG module are then automatically refined using Rietveld refinement within PXRDGen, ensuring optimal alignment between the predicted crystal structure and the PXRD data.

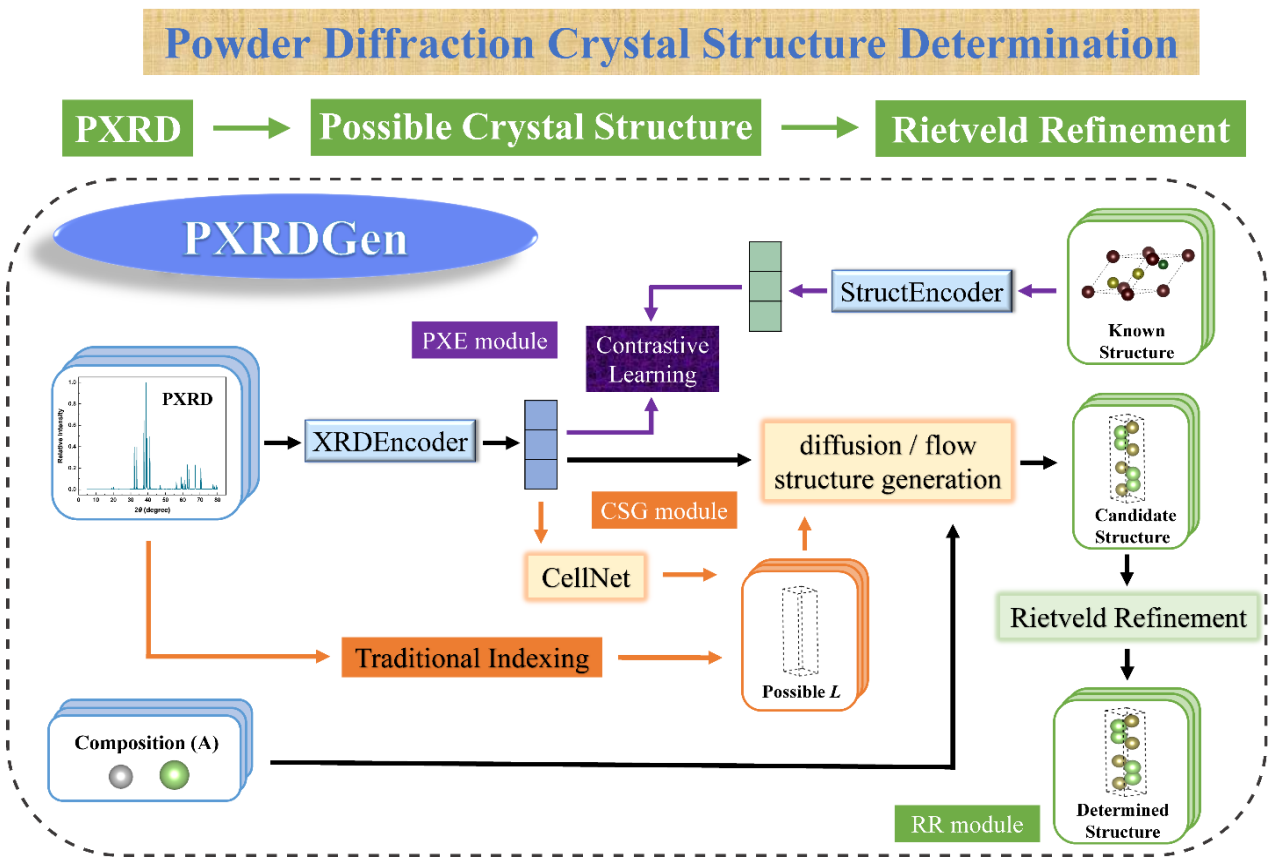

**Figure 1. Overview of PXRDGen.** The purple arrow indicates the pre-training of the XRD encoder (PXE module). The orange blocks represent the crystal structure generation (CSG module) based on PXRD data and composition. The green section denotes the Rietveld refinement (RR module).

## 2.2 Stable Structure Dataset and Evaluation Methods

PXRDGen is designed to solve experimentally stable inorganic compounds. The dataset used for inorganic crystal structures comes from the Materials Project (MP)[36], specifically the MP-20 dataset, which predominantly consist of experimentally stable structures. The dataset features crystal structures in their primitive cells, with no more than 20 atoms per cell. The corresponding PXRD patterns are calculated using GSASII[37] with default instrument parameters, employing both Cu Kα1 and Kα2 radiation. Each PXRD pattern is simulated over a $2\theta$ range of 5° to 80° with a step size of 0.01°, and normalized by maximum peak intensity. The MP-20 dataset comprises 45,229 structures, split into 60% training data, 20% validation data, and 20% test data. Besides, the automatic Rietveld refinement is applied to the predicted structures using GSASII scriptable.

To evaluate the outcomes, the resolved and refined structures are compared with the ground truth using the StructureMatcher in Pymatgen[38], with settings of *stol* = 0.5, *angle_tol* = 10, and *ltol* = 0.3. The match rate and RMSE are provided to assess the quality of the predicted inorganic crystal structures. Notably, a noticeable portion of compounds in MP-20 (8.93%) is invalid and not matchable, so that the actual match rate is underestimated without removing the invalid entries. To provide a more accurate assessment, the match rate only for the valid entries in MP-20 is also reported.

## 2.3 Pre-training for XRD encoder

Pre-training is a well-established method for enhancing multimodal learning capabilities. To align the latent space of PXRD patterns with crystal structures, the contrastive learning method[28] has been applied. Here we design two encoders to extract the latent features from PXRD patterns, based on convolutional neural network (CNN) and transformer (T) architecture respectively. Besides, we have designed another encoder for crystal structure information extraction, which satisfies periodic E(3) invariant as described in DiffCSP. During the process of pre-training module, the InfoNCE loss is used as follows,

$$\text{Loss} = -\sum_{i=1}^{N} \log \frac{e^{sim(P_i, C_i)/t}}{\sum_{j=1}^{N} e^{sim(P_i, C_j)/t}} - \sum_{i=1}^{N} \log \frac{e^{sim(C_i, P_i)/t}}{\sum_{j=1}^{N} e^{sim(C_i, P_j)/t}}.$$

Here $P_i$ and $C_i$ refers to the extracted feature of PXRD and crystal structure in latent space, and $sim(P_i, C_i)$ calculates the cosine similarity between them. Besides the hyper-parameters t is named as temperature coefficients, which is vital to the outcomes of pre-training.

The heatmap of similarity metrics of random 100 samples in test dataset of MP-20 based on Transformer and CNN are depicted in Figure 2(a, b) respectively, which represents the goodness of alignment in latent space between PXRD and structure. More details of heatmap can be seen in Supplementary Figure S1. Figure 2(c) displays the top-k retrieval hit rate with the changing temperature coefficients, showing that the hit rate increases significantly with the decreasing temperature coefficients, especially for the XRDEncoder-T. From all these pre-training outcomes, it can be concluded that encoders based on Transformer perform better than CNN, and the top-10 hit rate reaches 92.42% in XRDEncoder-T (t=0.05), while the top-10 hit rate of XRDEncoder-CNN (t=0.05) is much lower, reaching 33.57%.

**2.4 Optimized strategy in the CSG module**

The CSG module in PXRDGen uses information extracted from different XRD encoders to condition the generation of crystal structures, so the design choices of PXRD encoder and CSG modules can significantly impact performance. As shown in Figure 2(d), in the case of the diffusion-based CSG network, various XRD encoders have highly different one-sample match rate. Among the various encoders, the CNN-based XRD encoder (orange columns) generally outperforms the Transformer-based encoder (green columns), in contrast with the results of contrastive learning. Specifically, the CNN-based encoder with unfixed pre-trained parameters demonstrates the strongest ability to generate accurate structures. For a benchmark, structural features directly extracted from the target structures (ground truth) are introduced into the diffusion model, achieving a high one sample match rate of 67.74%. As shown in Figure 2(e), the result obtained with the best CNN-based encoder already approaches this ground truth results - achieving one sample match rate up to 66.01%, underscoring the high effectiveness of the CSG module under the condition of PXRD features.

In the case of the Transformer-based XRD encoder (Figure 2d), fixing the pre-trained parameters (striped green columns) consistently leads to better performance, yielding higher match rate and lower RMSE than when the parameters are unfixed during subsequent training (blank green columns). This trend holds true regardless of the pre-training quality. A closer look at the light green and bright green columns reveals that higher pre-training quality led to superior structure generation, with the fixed, well-pretrained parameters (bright-green-striped columns) achieving the best results: 60.09% match rate for one sample. In the case of the CNN-based XRD encoder (Figure 2d), however, unfixing the pre-trained parameters (blank orange columns) results in a significant performance boost than these with fixed pre-trained parameter (striped orange columns). Moreover, the quality of pre-training appears to be less critical for the CNN-based encoder, as both well-pretrained and less-pretrained models deliver comparable performance. However, these pretrained models significantly outperform encoders without pretraining, achieving the best results with a 66.01% match rate for a single sample.

Due to the random nature of the diffusion process, multiple stable crystal structures can be generated by the CSG module to further increase the match rate. As shown in Figure 2(d), the match rates of different XRD encoders all increase significantly as the number of candidates increases from 1 to 20. Moreover, the RMSE value tends to decrease with multiple generations, with an average RMSE around 0.05, indicating that PXRDGen can generate highly precise crystal structures for experimentally stable compounds. The lowest RMSE, approximately 0.04, was achieved using the two best CNN-based encoders, which also corresponded to the highest 20-sample match rates of 83.87% and 84.04%, respectively.

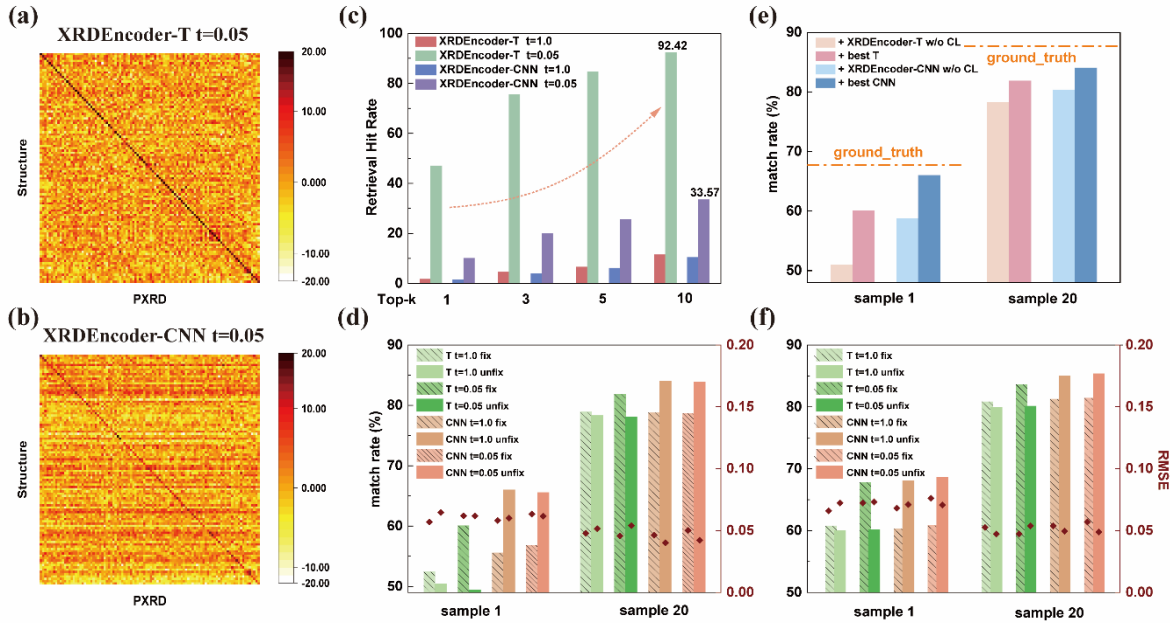

**Figure 2. Results of the PXRDGen model on the MP-20 dataset.** (a, b) Heatmaps of similarity metrics for 100 random samples from the test dataset in the PXE module, using XRDEncoder-T (t=0.05) and XRDEncoder-CNN (t=0.05), respectively. (c) Top-k retrieval hit rates for various XRD encoders with different temperature coefficients in the PXE module. (d) Results of the diffusion-based CSG module, showing the match rate (columns) and RMSE (wine-red diamond markers). (e) Summary of the diffusion-based CSG module with various XRD encoders. (f) Results of the flow-based CSG module. Additional details are provided in Supplementary Tables S1 and S2.

**2.5 Highly effective and precise structure solution based on flow model**

While diffusion models have achieved some success in generating stable materials, flow-based models are better suited for crystal structure prediction due to the more flexible and direct generation paths, seen in Figure 3(a). In PXRDGen, we also designed flow-based CSG module for generating crystal structures conditioned on PXRD data and compositions. Training flow-based models involves directly maximizing data likelihood, making them more efficient and superior to diffusion models, with higher match rates and significantly faster generation speeds. As shown in Figure 3(c), the flow model achieves a one-sample match rate of 69% that converged in just 50 generative steps — compared to 66% in the best diffusion model, which requires 1,000 steps. Additionally, the RMSE of structures generated with 200 steps in the flow model is comparable to that of diffusion models requiring 1,000 steps (~0.06). Consequently, the average time to generate one stable structure candidate is reduced to approximately 1 second using the flow model — nearly 4 orders of magnitude faster than quantum mechanical computations and 5 times faster than diffusion models.

Figure 2(f) displays the response of flow model to various pre-training XRD encoders in structure generation, which is consistent with that of the diffusion model. For instance, in terms of XRDEncoder-T, the best results are obtained by fixing the optimized pre-trained parameters, result in 67.84% match rate in 1 sample and 83.63% in 20 samples, corresponding to 74.48% and 91.83% for valid compounds respectively. For XRDEncoder-CNN, by unfixing the pre-trained parameters, the CSP module reached 68.68% match rate in 1 sample and 85.37% in 20 samples, corresponding to 75.41% and 93.74% for valid compounds respectively. As illustrated in Figure 3(b), under the cases of various XRD encoders, the flow model is consistently superior to the diffusion model in match rate at much lower computational costs.

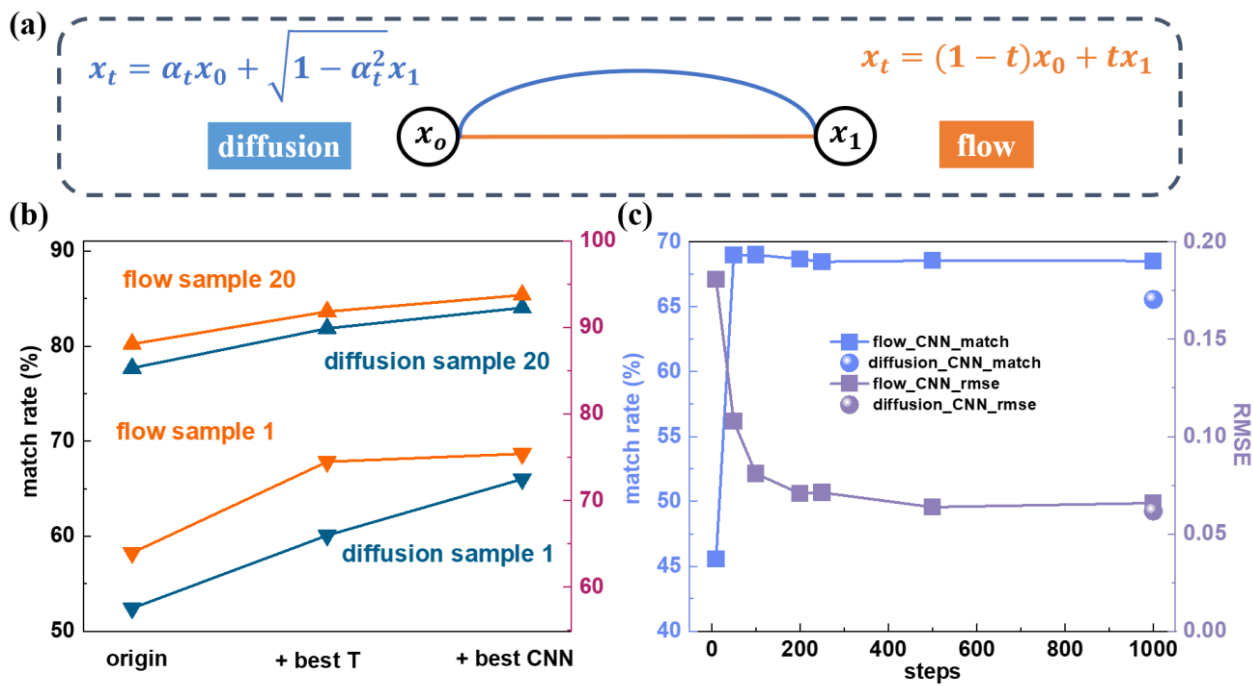

**Figure 3. Comparison between flow-based and diffusion-based CSG modules on MP-20.** (a) Illustration of the different generation paths in diffusion and flow models. (b) Match rates of the flow and diffusion models for 1 and 20 sample generations; the left side shows the match rate for all compounds, while the right side indicates the match rate for valid compounds. (c) Comparison of match rates and RMSEs for candidate structures (1 sample) generated by the flow and diffusion models at different generative steps.

## 2.6 Better performance with given *L* in solving structures

In conventional Crystal Structure Prediction (CSP) tasks, the stable 3D structure is predicted based solely on its composition. The lattice parameters (*L*) cannot be explicitly determined and should be better generated jointly with fractional coordinates ($F_i$). PXRD data allows for independent determination of *L* based on peak positions ($2\theta$). This allows for the correct *L* to be supplied as an additional condition when predicting $F_i$, as illustrated in Figure 4(a). For instance, the flow model combined with XRDEncoder-CNN (flow-CNN) has demonstrated optimal performance in the joint generation of both $L$ and $F_i$. By providing the true $L$ that can be routinely obtained through indexing software, the match rate of flow-CNN improves significantly. As shown in Figure 4(b), the one-sample match rate increases from 68.68% to 75.32%, enabling a considerable fraction of previously mismatched structures to be correctly resolved. For example, in the case of $NaTiVS_4$, the flow-CNN module initially failed to predict the correct structure, with the $L$ significantly changes at different generation steps. Moreover, when conditioned with the correct $L$, it successfully generated the correct structure with RMSE below 0.04, as depicted in Figure 4(c). The one-sample match rate of 75.32% is exceptionally high, corresponding to 82.70% for valid compounds. With multiple sampling, the 20-sample match rate reaches 88.08% for all compounds and 96.71% for valid compounds.

In PXRDGen, the $L$ obtained by indexing PXRD patterns can be directly utilized. If indexing results are unavailable, PXRDGen provides a neural network, called CellNet, trained to extract the $L$ directly from PXRD patterns, as shown in Figure 4(a) and Supplementary Figure S2(a). CellNet can generate $L$ from a PXRD pattern within milliseconds, so that multiple $L$ can be generated to enhance the precision of the cell parameters determination. The ability of CellNet is detailed in Supplementary Table S3. It's shown that the Mean Absolute Percentage Error (MAPE) of the best cell parameters is around 1% for 1000 samples. The optimal cell can then be selected and used as a condition in the flow-CNN module to generate crystal structures. As shown in Figure 4(d), incorporating the optimal $L$ in the structure generation process significantly improves performance. The orange line represents the results by selecting the best $L$ with the lowest MAPE values from multiple samples ($N$), which already outperforms the flow-CNN module at $N = 20$ and approaches the upper limit value of ground truth $L$ at $N = 1000$. The match rate increases with the number of candidates of $L$, indicating that multiple generations are effective for achieving precise structure predictions.

While we have demonstrated that CellNet approaches the limits of providing ground truth values for $L$, practical algorithms still need to be developed to select the best $L$ from multiple candidates based on diffraction positions. The procedure is in fact challenging, as the existence of high symmetric lattice and extinction effects can lead to very different number of diffraction peaks of target and predicted $L$, making the calculation of similarity between the $L$ difficult. Dynamic Time Warping (DTW) is successfully used in describing the similarity of different lengths of sequences, which can be used to screen out the matched $L$ since a slight deviation of symmetry leads to drastic change in diffraction patterns, as demonstrated in Supplementary Figure S2(b). Using the FastDTW[39] screen algorism, the incorporation of CellNet outperforms the flow-CNN module at $N = 100$ and the one-sample match rate reaches 70.39% at $N = 1000$.

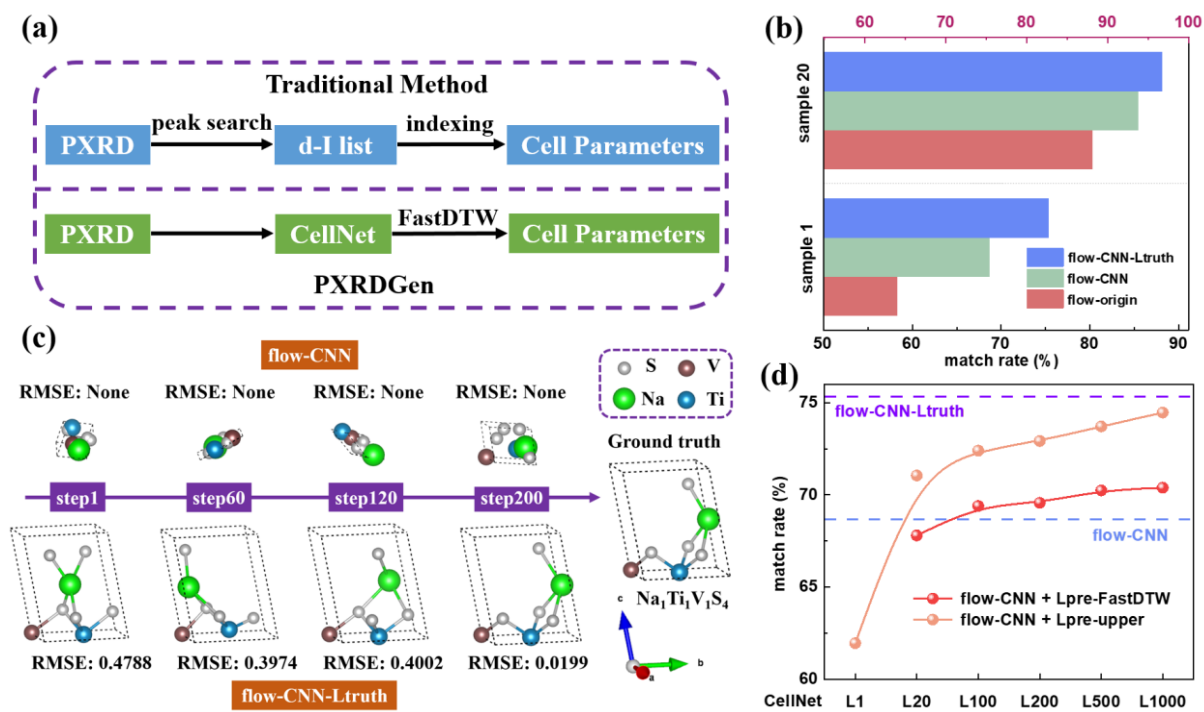

**Figure 4. Structure solution using PXRDGen with lattice information.** (a) Schematic diagram comparing the traditional method and CellNet for obtaining lattice parameters ($L$). (b) Performance of the flow-CNN module when provided with correct lattices versus not. The bottom section represents the match rate for all compounds, while the top section refers to the match rate for valid compounds. (c) The generation process of $NaTiVS_4$ with $L$ jointly generated with $F_i$ and correct $L$. (d) One-sample match rate of the flow-CNN module when using $L$ predicted by CellNet. $L_N$ refers to selecting the best lattice from N predicted lattices by CellNet. The orange line represents the upper capability of CellNet combined with flow-CNN, where the best $L$ is chosen by directly comparing it to the target $L$. The red line shows the results using the best $L$ selected through FastDTW. Additional results are provided in Supplementary Table S4.

## 2.7 Application of PXRDGen in Solving Challenging Crystal Structures

Having demonstrated that PXRDGen can solve crystal structures with unprecedented accuracy, we also show that this AI-based approach offers a superior understanding of stable structures, surpassing the detection limits of traditional XRD techniques. While XRD is a powerful tool for determining crystal structures by mapping electron clouds, it has notable limitations, particularly in locating light elements amidst heavier ones and distinguishing between elements with similar atomic numbers. This challenge is especially pronounced when trying to detect hydrogen, which, with only one electron, is often invisible in XRD data — even in some cases in single-crystal diffraction. Accurately locating hydrogen and differentiating neighboring elements in a crystal structure often requires more costly neutron diffraction techniques. Additionally, the serious overlapping of peaks in PXRD poses an additionally challenge for conventional structure solution methods.

The AI-based PXRDGen model addresses these challenges by leveraging its understanding of stable crystal structures beyond the resolution capabilities of conventional PXRD. For instance, in the case of $Zr_2Ni_2H_6$ (Figure 5a), conventional PXRD can reliably determine the positions of the heavy elements Zr and Ni but leaves the hydrogen atoms unresolved[40] — a common issue in hydrogen-containing structures. However, PXRDGen successfully resolved the complete structure, with the positions of all hydrogen atoms perfectly aligning with those determined by neutron diffraction (RMSE < 0.02). Similarly, for $Li_6Au_2S_4$ (Figure 5b), the lithium atom locations identified by PXRDGen matched precisely with results from single-crystal diffraction[41]. Furthermore, as shown in Figure 5(c), PXRDGen can generate the correct structure of $Sr_2H_8O_6$ which is similar to experimental structure by neutron diffraction[42] while the SCXRD[43] cannot give the position of all the H atoms.

PXRDGen significantly enhances the differentiation of elements with similar atomic numbers, a task that typically requires neutron or transmission electron diffraction rather than PXRD. For instance, in $Mn_2Fe_2As_2$ (Figure 5d), where Mn and Fe have similar atomic numbers, neutron diffraction is usually necessary to distinguish between them[44]. However, PXRDGen accurately identified the correct element assignments. This capability extends to complex structures like $Sr_3Fe_2Cu_2Se_2O_5$ (Figure 5e), where differentiating Fe from Cu is challenging due to their similar coordination environments with Se and O[45]. PXRDGen effectively positioned Fe and Cu with a high precision (RMSE = 0.025), a task typically requiring high expertise in coordination chemistry. Additionally, PXRDGen excels in cases with overlapping peaks, which poses a significant challenge in traditional PXRD. For example, in $Y_6Pd_1Br_{10}$ (Supplementary Figure S3), where the PXRD pattern is complicated by peak overlap, PXRDGen successfully generated the correct structure with an RMSE of 0.015, a task that have required SCXRD in experiments[46].

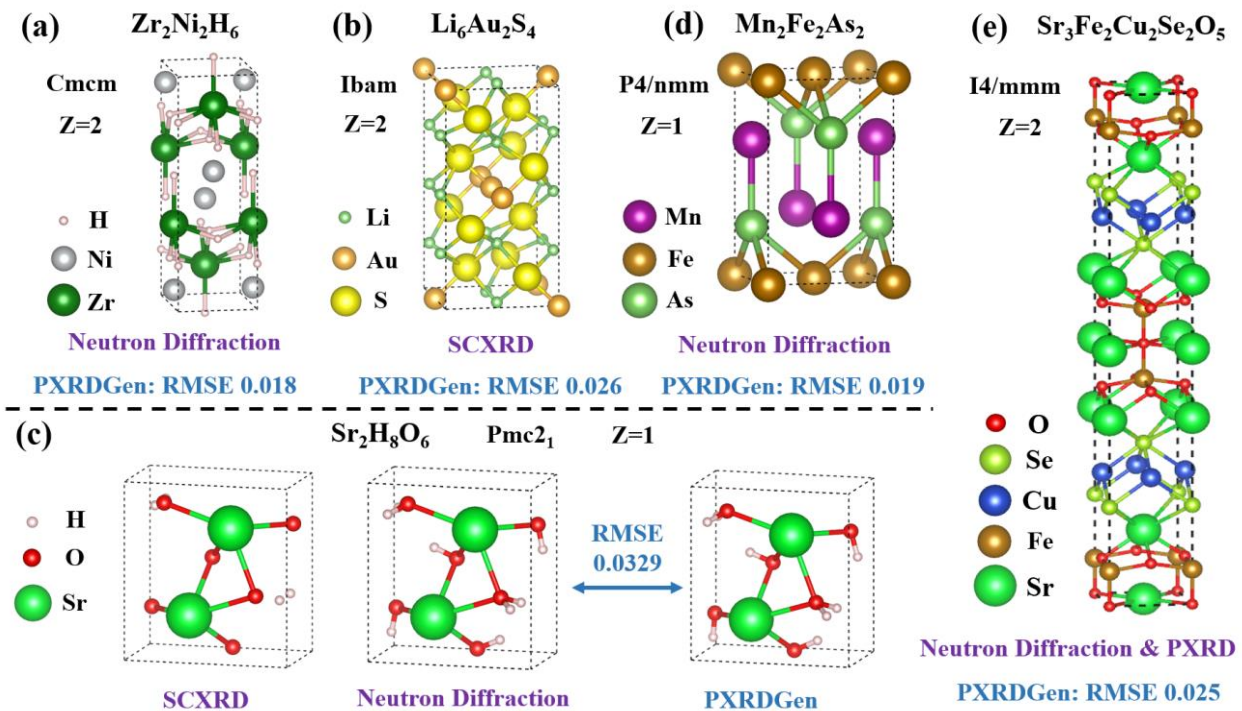

**Figure 5. Examples of PXRDGen in Solving Challenging Structures.** Application of PXRDGen in locating light elements amidst heavier ones is demonstrated in (a) $Zr_2Ni_2H_6$, (b) $Li_6Au_2S_4$ and (c) $Sr_2H_4O_6$. PXRDGen's ability to distinguish between elements with similar atomic numbers is shown in (d) $Mn_2Fe_2As_2$ and (e) $Sr_3Fe_2Cu_2Se_2O_5$. In all cases, the Z represents that number of chemical formulas in a conventional cell.

## 2.8 Rietveld Refinement in crystal structures determination

The initial structural model obtained from structure solution generally requires Rietveld refinement to improve its fit to experimental data. However, due to the nature of the nonlinear least-squares algorithm at the core of Rietveld refinement, the process is highly sensitive to the accuracy of the initial model. By leveraging the precise structure models generated by the CSP module, we have developed an automated Rietveld refinement module within PXRDGen. To assess its effectiveness, we have tested the predicted structures from the flow-CNN-Ltruth module in the MP-20 dataset. Out of 9,046 materials in the test set, 8,238 are valid, with PXRDGen successfully solving the structure for 7,946 materials (96.5%) for the 20-sample approach. Among these, 7,304 structures were successfully refined using automatic Rietveld refinement. Details of the outcomes are provided in Supplementary file.

As shown in Figure 6(a), the RMSE distribution of 7,304 materials decreased significantly by an order of magnitude following Rietveld refinement. Figure 6(b) categorizes the predicted materials into six regions based on RMSE, with each region indicating that the predicted RMSE is smaller than the corresponding threshold. It is evident that for the vast majority of predicted structures, the RMSE is below 0.05, and this is further reduced after Rietveld refinement. However, the refinement process becomes increasingly challenging as the initial RMSE of the predicted structures rises. Specifically, Rietveld refinement struggles when the predicted RMSE exceeds 0.3 and largely fails when it surpasses 0.4, underscoring the importance of initial model accuracy for successful automatic Rietveld refinements.

Figure 6(c, d) highlight two typical cases with various initial RMSE values and structural complexity. For structures without light elements, Rietveld refinement generally reduces the RMSE significantly, bringing the predicted structure into closer alignment with the PXRD data. This module not only enhances the match between the predicted structure and experimental data but also provides a reliable benchmark for selecting the best candidate structure from multiple generated options. While Rietveld refinement substantially reduces the RMSE for most crystal structures, a small fraction of structures (<5%) showed increased RMSE after refinement (Supplementary Figure S4). In these cases, the CSG module may more effectively capture stable structures compared to Rietveld refinement alone, which depends solely on X-ray diffraction resolution and does not account for the rationality of the crystal structure. Integrating AI capabilities into Rietveld refinement in the future should help address this issue.

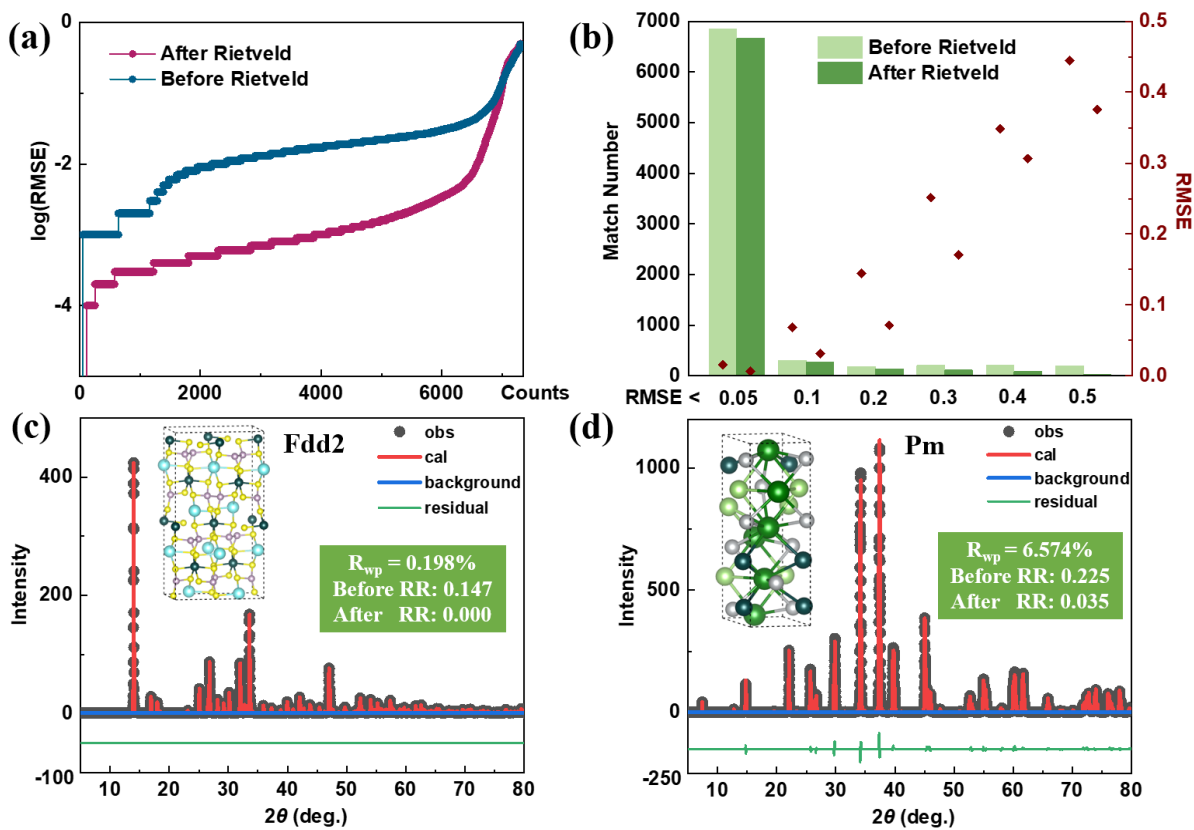

**Figure 6. Rietveld refinements after crystal structure prediction on MP-20.** (a) The distribution of RMSE before and after Rietveld refinement. (b) Changes in matching number in various RMSE regions of predicted structures. Example of (c) $Na_2Al_2P_4S_{12}$ and (d) $Ce_6Al_3Ga_3Ni_6$ after Rietveld refinement, constrained with space group. $R_{wp}$ is the weighted factor to present the quality of the Rietveld fit. Here the $R_{wp}$ is smaller than experiment because of the simulated PXRD pattern.

## III. Conclusion

In conclusion, we introduce a novel AI-based approach for automatically and effectively solving and refining crystal structures from powder diffraction patterns, successfully applying it to thousands of compounds with diverse symmetries and elemental compositions. The CNN-based XRD encoder is highly optimized, closely approximating ground truth values. The extraction and conditioning of PXRD features and lattice parameters have significantly enhanced the structure-solving capability of PXRDGen, powered by Riemannian flow matching module. This approach achieved an unprecedented high match rate of up to 82% for a single candidate and 97% for 20 candidates, for valid compounds in the MP-20 database. PXRDGen also excels in resolving challenging structures containing light atoms and elements with similar atomic numbers, thanks to its deep understanding of stable structures. Additionally, the precise atomic structures generated are automatically refined using Rietveld refinement, eliminating the need for time-consuming manual adjustments and surpassing the limitations of human chemical intuition and PXRD resolution. This paradigm represents a significant advancement in the automation of PXRD analysis and crystal structure determination.

**Methods and Codes**

Details of neural network are provided in Supplementary File. And our codes are available once the article is published.

**ACKNOWLEDGMENTS**

This work is financially supported by the National Natural Science Foundation of China (Grant No. 52272268), the Key Research Program of Frontier Sciences, CAS (Grant No. QYZDJ-SSWSLH013), the Informatization Plan of Chinese Academy of Sciences (Grant No. CAS-WX2021SF-0102), the National Natural Science Foundation of China (Grant No. 62376276), Beijing Nova Program (Grant No. 20230484278), the National Natural Science Foundation of China (Grant No. 61925601).